\documentclass[iop, revtex4, twocolappendix, numberedappendix]{emulateapj}
\usepackage[pass]{geometry}
\usepackage{amsmath}
\usepackage{setspace} 
\usepackage{graphicx}
\usepackage[english]{babel}
\usepackage[utf8]{inputenc} 
\usepackage{slashed}
\usepackage[backref,breaklinks,colorlinks,citecolor=blue]{hyperref}
\usepackage{epstopdf}
\usepackage[all]{hypcap}
\shorttitle{Gender Bias in Astronomy}
\shortauthors{Caplar, Tacchella \& Birrer}

\usepackage{color}

\begin{document}

\title{Quantitative Evaluation of Gender Bias in Astronomical Publications \\ from Citation Counts }

\author{Neven Caplar\altaffilmark{1}, Sandro Tacchella\altaffilmark{2} \& Simon Birrer\altaffilmark{3} \\{Institute for Astronomy, Department of Physics, ETH Zurich, CH-8093 Zurich, Switzerland}}
	    
\altaffiltext{1}{\href{mailto:neven.caplar@phys.ethz.ch}{neven.caplar@phys.ethz.ch}}
\altaffiltext{2}{\href{mailto:sandro.tacchella@phys.ethz.ch}{sandro.tacchella@phys.ethz.ch}}
\altaffiltext{3}{\href{mailto:simon.birrer@phys.ethz.ch}{simon.birrer@phys.ethz.ch}}

\submitted{\today}

\begin{abstract}
We analyze the role of first (leading) author gender on the number of citations that a paper receives, on the publishing frequency and on the self-citing tendency. We consider a complete sample of over 200,000 publications from 1950 to 2015 from five major astronomy journals. We determine the gender of the first author for over 70\% of all publications. The fraction of papers which have a female first author has increased from less than 5\% in the 1960s to about 25\% today. We find that the increase of the fraction of papers authored by females is slowest in the most prestigious journals such as Science and Nature. Furthermore, female authors write $19\pm7\%$ fewer papers in seven years following their first paper than their male colleagues. At all times papers with male first authors receive more citations than papers with female first authors. This difference has been decreasing with time and amounts to $\sim$6\% measured over the last 30 years. To account for the fact that the properties of female and male first author papers differ intrinsically, we use a random forest algorithm to control for the non-gender specific properties of these papers which include seniority of the first author, number of references, total number of authors, year of publication, publication journal, field of study and region of the first author's institution. We show that papers authored by females receive 10.4$\pm$0.9\% fewer citations than what would be expected if the papers with the same non-gender specific properties were written by the male authors. Finally, we also find that female authors in our sample tend to self-cite more, but that this effect disappears when controlled for non-gender specific variables. \\
\end{abstract}


\maketitle

\section{Introduction}

	Gender inequality and biases seem to be persistent in the scientific community. Even though the number of doctorate degrees awarded to women is constantly increasing, women still tend to be underrepresented in faculty positions \citep{NSF15}. Numerous studies have shown that both male and female referees consistently give higher scores to identical work done by males than females (e.g., \citealp{W97}, \citealp{M12}). As an example of bias in publishing, the study by \citet{B08} showed that the number of female authors increased significantly after a journal in the field of ecology switched to the double-blind refereeing system in which the names of the authors are kept hidden from the reviewers.\\
	
    The recent growth of big databases enables more systematic statistical investigation into the role of gender on the publishing and awarding mechanisms in the scientific community. \citet{C12} deduced that female authors tend to be underrepresented in the prestigious publications; for example, female authors have contributed only 3.8\% of earth and environmental science articles for Nature News \& Views even though they represent approximately 20\% of scientists in the field. The same conclusion was reached by \citet{W13} who conducted a large multi-field analysis and found much of the disparity between male and female authors was due to lack of females who are first authors of prestigious papers. The same group has also found that men tend to self-cite their work more \citep{K16}. For the domain of engineering, \citet{G15} has recently showed around 10\% bias in the number of citations.\\
    
  Focusing on astronomy, \citet{D14} has studied gender balance at the 223rd meeting of the American Astronomical Society and found that even though the gender ratio of speakers mirrors that of conference attendees, women asked fewer questions than their male peers. A similar conclusion was reached by \citet{P15} who studied patterns at the National Astronomy Meeting 2014 of UK astronomers. A study by \cite{R14} on the success of proposals for time on the Hubble Space Telescope concluded that proposals with a female principal investigator are less likely to succeed than proposals with a male principal investigator. They also found that the success rates by males and females for more recent graduates (Ph.D. since 2000) are more comparable to each other. Similar disparity between genders was also recently reported for time allocation at European Southern Observatory telescopes \citep{P16}. Although these difference are observed in the conference settings and in the proposal success rate, no study has investigated possible differences in the number of citations between the genders.  \\

	 Spurred by these findings, we wish to measure the role of gender on the number of citations that papers receive in astronomy. Throughout the study we assume that male and female authors should receive the same number of citations for papers that have the same non-gender properties. Any difference in the citation counts between female and male first author papers with matched non-gender properties will be labeled ``gender bias''. The main goal of this paper is to quantify this gender bias.  \\ 
     
     Furthermore, we also investigate the publishing frequency and the self-citing tendency of the authors. We chose to analyze astronomy since it is our own field of research, and since a homogeneous data set for this field can be constructed. We use a large data set in order to be able to control for spurious dependencies of the number of citations on non-gender specific variables and to determine which factors are the drivers of the possible differences between male and female first author papers. \\  
        
    We often use the syntax ``female authors'' and ``male authors''. Firstly, it is to be understood that when we use this syntax we actually refer to the first authors of the papers. We are primarily interested in the gender of the first author of the paper since in astronomy the first author of the paper is the principle investigator. Secondly, we wish to make clear that our results are only understood within the constraints of our analysis. We are not able to determine the gender of the authors of each paper with absolute certainty and there may be some biases in our estimation of gender, which we discuss in Section~\ref{sec:discussion}. For all practical purposes, the phrases ``female authors'' and ``male authors'' are to be understood as ``first authors that we deduced to be female in this analysis'' and ``first authors that we deduced to be male in this analysis'', respectively. \\    
    
	The structure of the paper is as follows. In Section~\ref{sec:data} we describe our compilation and reduction of the data. In Section~\ref{sec:SampleChar} we present some characteristics of the sample and the differences between the male and female authored papers. In Section \ref{sec:simple_bias} we present the difference between the mean number of citations gathered by male and female first author publications and discuss how seniority of authors affects these results. In Section \ref{sec:ml_bias} we aim to isolate the effect of gender from the other non-gender specific variables using machine-learning techniques. Section \ref{sec:self_citation} discusses the self-citation dependence on gender. Section \ref{sec:discussion} contains discussion of the possible caveats and we conclude in Section~\ref{sec:conclusion}. In the Appendix we expand on several additional properties of our sample.\\

\section{Data}\label{sec:data}

	In this section, we describe the data compilation and cleaning processes we used in order to produce our final dataset. 

\subsection{Data sources}

	To get the list of all published papers in the field of astronomy we downloaded from the SAO/NASA Astrophysics Data System (ADS)\footnote{\url{http://adswww.harvard.edu/}} all of the entries available in the database ``astronomy'' and published in one of five established journals (``Astronomy \& Astrophysics'' (AA), ``Astrophysical journal'' (APJ), ``Monthly Notices of Royal Astronomical Society'' (MNRAS), ``Nature'' (NAT) and ``Science'' (SCI)) from 1950 to 2015. We choose these five journals as they encompass the vast part of astronomical research today. Furthermore, they are well established journals with long historical records. SAO/NASA astronomy API service provides many types of metrics for each paper. Specifically, we chose to download the names of the authors and their institutions, the number of citations, the number of references, the publishing journal's name, abstract of the paper and the publishing year. All of the information was downloaded in a single effort in June 2016, so the number of citations for every paper reflects state of the metric at that point in time. \\ % \jwc{you should mention how complete the database is.  I'm not sure that it is complete for the early years.  Also, specify that these are English only.  I believe there are several older astronomy journals in other languages.}
    
    We augment the data with information available from the arXiv database\footnote{\url{https://arxiv.org/}}, for papers where such data exist. For each paper that is found in the arXiv database we record the designated field (``Astrophysics of Galaxies"; ``Cosmology and Nongalactic Astrophysics"; ``Earth and Planetary Astrophysics"; ``High Energy Astrophysical Phenomena"; ``Instrumentation and Methods for Astrophysics"; ``Solar and Stellar Astrophysics") and downloaded the *.tex source file when possible from the Amazon S3 server\footnote{\url{http://arxiv.org/help/bulk_data_s3}} in order to determine the length of paper as well as the number of equations and floats in the paper.

\setcounter{table}{0}
\renewcommand{\thetable}{1\Alph{table}}

\capstartfalse 
\begin{deluxetable*}{lccccccc}
\tabletypesize{\scriptsize}
\tablecolumns{11}
\tablewidth{0pt}
\tablecaption{Example of the data available (first 8 columns) \label{tab:bigtab}}
\tablehead{   % column headings
\colhead{Bibcode} & \colhead{First Author$^{1}$} & \colhead{First name} & \colhead{Gender} & \colhead{first publication year$^{2}$}  & \colhead{\# citations} & \colhead{\# references} & \colhead{\# authors}  }
\startdata
1978ApJ...222..745C&{Condon,  J. J.}&James&male&1973&19&22&2\\

1988ApJ...333..611W&{Wilson, Christine D.}&Christine&female&-99&18&14&5\\

1990MNRAS.246..565A&{Aspin,  C.}&Colin&male&1981&19&26&4\\

1990Natur.345...49T&{Torbett,  Michael V.}&Michael&male&1980&48&11&2\\

1992ApJ...392..760B&{Burrows,  Christopher J.}&Christopher&male&1991&37&7&3\\

1993A\&A...277..677M&{Meier,  R.}&Roland&male&1993&97&77&4\\

1996A\&A...309..171S&{Shibanov,  Y. A.}&Yurii&male&1992&42&18&2\\

1997A\&A...324L...5C&{Cambresy,  L.}&Laurent&male&1997&58&12&8\\

2002A\&A...381L..25M&{Meynet,  G.}&Georges&male&1985&82&31&2\\

2002MNRAS.329L..67B&{Ballantyne,  D. R.}&David&male&2000&31&29&3\\

2010ApJ...711.1310K&{Khatri,  Rishi}&Rishi&male&2010&3&37&2\\

2014ApJ...780..111H&{Heitmann,  Katrin}&Katrin&female&2006&63&57&5\\
...
\enddata
\tablenotetext{1}{Name of the first author as specified in the paper}
\tablenotetext{2}{Year in which the leading author of the paper in question published their first paper}

\end{deluxetable*}
\capstarttrue

\capstartfalse 
\begin{deluxetable*}{cccccccccc}
\tabletypesize{\scriptsize}
\tablecolumns{11}
\tablewidth{0pt}
\tablecaption{Example of the data available (continued, last 9 columns) \label{tab:bigtab2}}
\tablehead{   % column headings
\colhead{Region} & \colhead{Year} & \colhead{Journal} & \colhead{\# field$^{3}$} & \colhead{\# floats$^{4,5}$} & \colhead{\# equations } & \colhead{\# math inline } & \colhead{\# words} & \colhead{\# Bibcode of first publication}}
\startdata
NAMERICA&1978&APJ&3&-99&-99&-99&-99&1973ApJ...183.1075C\\

NAMERICA&1988&APJ&4&-99&-99&-99&-99&-99\\

OTHER&1990&MNRAS&4&-99&-99&-99&-99&1981MNRAS.194..283A\\

NAMERICA&1990&NAT&1&-99&-99&-99&-99&1980Natur.286..237T\\

NAMERICA&1992&APJ&6&-99&-99&-99&-99&1991ApJ...369L..21B\\

OTHER&1993&AA&4&-99&-99&-99&-99&1993A\&A...277..677M\\

OTHER&1996&AA&2&-99&-99&-99&-99&1992A\&A...266..313S\\

OTHER&1997&AA&4&-99&-99&-99&-99&1997A\&A...324L...5C\\

EUROPE&2002&AA&2&-99&-99&-99&-99&1985A\&A...150..163M\\

EUROPE&2002&MNRAS&5&-99&-99&-99&-99&2000ApJ...536..773B\\

NAMERICA&2010&APJ&3&8&10&160&2709&2010ApJ...711.1310K\\

NAMERICA&2014&APJ&3&17&14&502&11456&2006ApJ...642L..85H\\
...
\enddata
\tablenotetext{3}{1=``Earth and Planetary Astrophysics", 2=``Solar and Stellar Astrophysics", 3=``Astrophysics of galaxies", 4=``Cosmology and Extragalactic Astrophysics", 5=``High Energy Astrophysical Phenomena", 6=``Instrumentation and Method for Astrophysics" }
\tablenotetext{4}{floats include both figures and tables}
\tablenotetext{5}{with -99 we denote that there is no data available for this quantity }
%\tablecomments{TBD.}
\end{deluxetable*}
\capstarttrue
\begin{figure*}[htp]
    \centering
    \includegraphics[width=.99\textwidth]{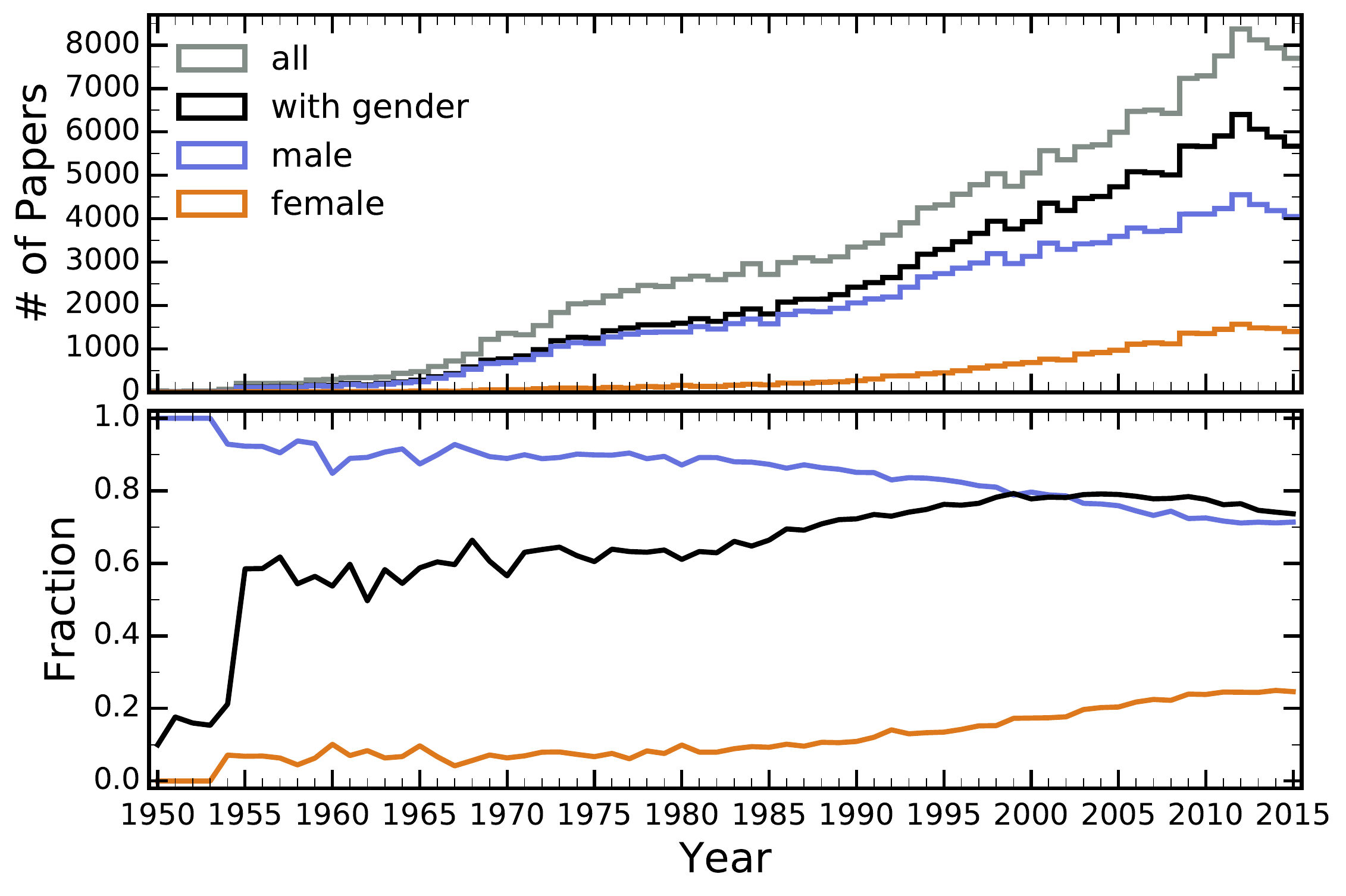}
    \caption{\textit{Top panel:} Number of papers published per year. The gray, black, blue, and orange histogram show the number of all papers (papers with more than zero citations and references), papers with determined first-author gender (sample used for analysis), papers with a male first author, and papers with a female first author, respectively. The overall number of papers published per year increased by more than one order of magnitude from the 1960s to 2015. \textit{Bottom panel:} Fraction of all papers for which we determined the gender (black), fraction of papers with gender for which first author is male (blue) and for which first author is female (orange), respectively. From 1980 onward, we determined the gender for 60-80\% of all papers. The fraction of female first author papers has steadily increased over the past 50 years.}
    \label{fig:sample}
\end{figure*}
\subsection{Adding paper-specific information}

	In this section we describe how we determine for each paper its length and its subfield. When the *.tex files are available we run the tool TeXcount\footnote{\url{http://app.uio.no/ifi/texcount/}} with default settings to obtain the number of words, floats, equations and mathematical expressions embedded in the text of each paper. For some papers the tool fails or measures very small number of words in the paper ($<500$) due to multiple *.tex files associated with the single paper. In these cases, we ignore these measurements in further analysis.\\ 

    In order to estimate the topic of the paper for which arXiv classification is not available, we train a random-forest algorithm on the sample of papers for which both field classification and their abstract are available. We are able to achieve high accuracy of classification; we find about 80\% of papers are being correctly classified. Reassuringly, the misclassification is often between similar categories, e.g., between ``Cosmology and Nongalactic Astrophysic" and ``Astrophysics of Galaxies" or between ``Earth and Planetary Astrophysics" and  ``Solar and Stellar Astrophysics". If we exclude these similar misclassifications we find that the accuracy rises to around 90\%. We then use this algorithm on all other papers in order to assign them their field of research.

\subsection{Adding author-specific information: institution}

	In order to simplify and categorize the institutional information for each paper we determine the country of the institution of the first author. In total, 85\% of papers include institutional information.\\ 
    
    We developed a list of about 100 keywords for which individual appearance in the affiliation string uniquely determines the country of origin. This list includes different spellings of country names, country codes, US state names and abbreviations, and university and research institution names. Linking the affiliation strings to this list enables us to assign 97\% of papers with affiliations uniquely to a country.\\
    
	To simplify this information, we assign the institutions to three categories: North America, Europe, and Other. We experimented with different classifications and found that these have minimal effect on our conclusions.
    
   \begin{figure*}[!htp]
    \centering
    \includegraphics[width=.99\textwidth]{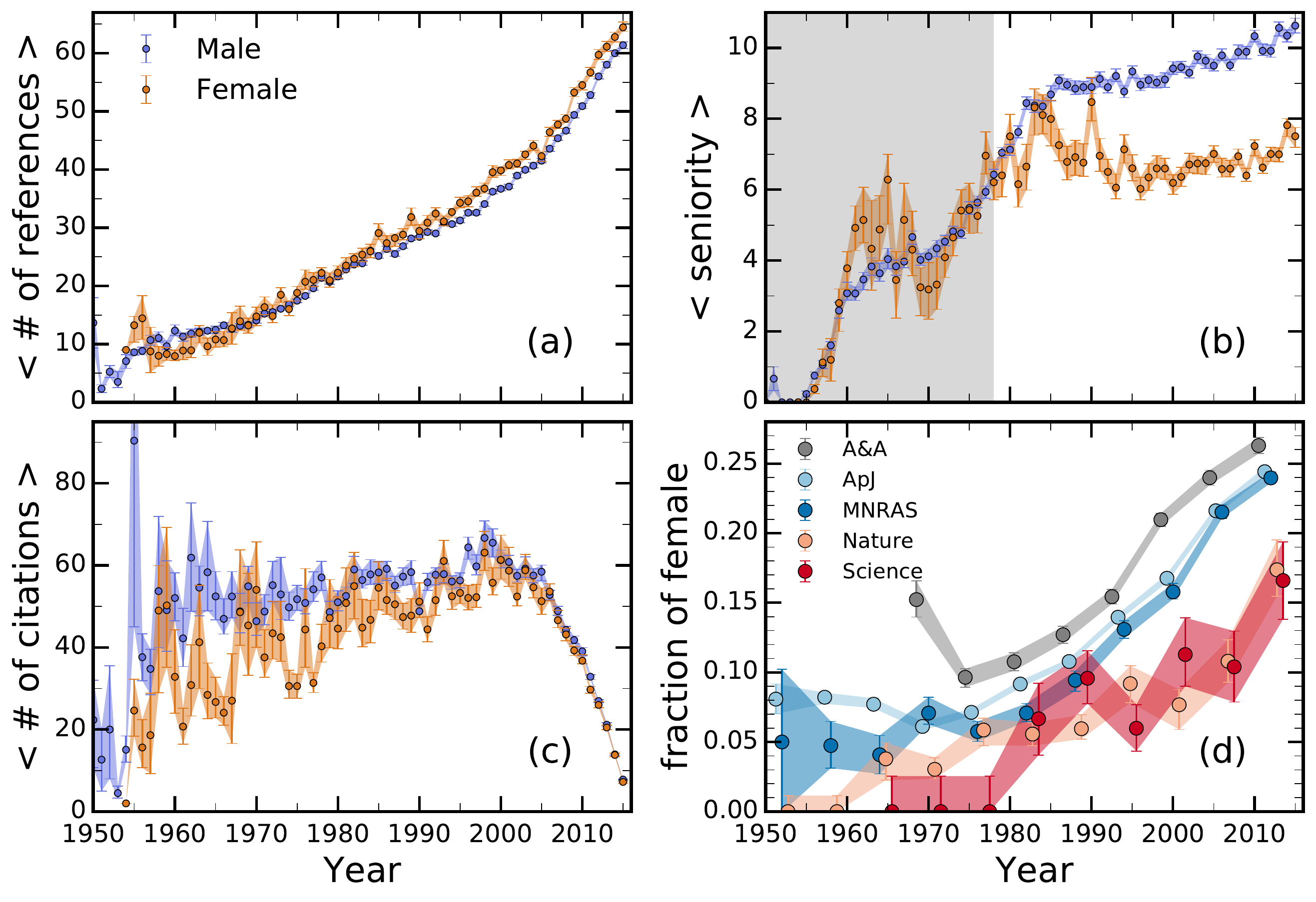}
    \caption{Characteristics of our sample. (\textit{a}) Mean number of references in the female (orange) and male (blue) first author papers. Females authored papers tend to have more references. (\textit{b}) Mean seniority of papers authored by females and males. The gray shaded regions indicate where our seniority evaluation is incomplete (see Appendix \label{app:completness} for more details). (\textit{c}) Mean number of citations. Papers with a female first-author have on average fewer citations. (\textit{d})  Fraction of female first author papers in different journals. Points are offset from each other in the horizontal axis for clarity. Data is stacked in 5-year intervals to shows the result more clearly. Female first authors are clearly underrepresented in high-impact journals such as ``Science'' and ``Nature''. The errors in all panels are obtained by bootstrapping and they denote the error on the mean of the quantity in question.}
    \label{fig:sample_characteristics}
\end{figure*}
    
\subsection{Adding author-specific information: gender and seniority} 

	Determining the first author's gender is complex because many authors publish using their initials instead of their full first names. We partially mitigate this problem by matching first and last names with initials of all authors from the dataset of all papers. In this way, we are able to determine the first name of an author even if they provided only initials in the particular paper, but used their first name at least once during their publishing career. We took special care in order to ensure bijection between author information with initials only and corresponding author information with a full first name as well. In many cases, the second and third first name (middle names) help to identify the unique full name provided by the initials. Thanks to this methodology, we are able to uniquely identify different authors in the entire dataset and their reappearance. We use the year of an author's first first-author paper as the baseline to define the seniority of an author. We define the seniority of an author as the number of years that have passed since their initial first-author publication. In the cases where the exact first paper of the author can not be identified due to possible confusion between the authors with same initials we do not assign a seniority to such an author. In addition, we looked for authors that have changed their last name by looking for authors with last names that are parts of other last names, while having the same first name. All possible cases have been individually checked whether indeed a change of last name is present. With this procedure we are able to recover full records for authors that have changed their surnames during their publishing carrier (e.g. due to marriage). \\

	After determining the full first name, we match the name to three different databases to determine the gender. Firstly, we look the name up with SexMachine\footnote{\url{https://pypi.python.org/pypi/SexMachine/}}, a python module. This database consists of 40,000 names from a wide geographical origin which have been classified by native speakers. %The gender categories are: androgynous, male, female, mostly male, and mostly female. 
Secondly, we search for gender in the data available from the United States Social Security Administration and the UK Office of National Statistics, which track the gender of all children born in these countries\footnote{\url{https://github.com/OpenGenderTracking/globalnamedata}}. %This list is highly complete as it tracks all children names which have been given more than 5 times in each year. 
It consists of approximately 100,000 names, but it does not have the geographical width as the first database. Finally, if the name is not found in those lists, we look the name up in Gender API\footnote{\url{https://gender-api.com/}}, which includes nearly 2,000,000 names. If a given first name consists of several names, we have checked the gender for all of the names, and weighted the final gender assignment accordingly.

\subsection{Cleaning and finalizing the dataset}

	The last step in our data processing is to remove the parts of the dataset which we judge to contain spurious information or for which we have incomplete information. In total, we have downloaded 208,577 entries from ADS. We remove 58,836 entries (about 28\%) from this initial dataset, giving us a final dataset with 149,741 papers. The following ADS entries are removed:
    $(i)$ entries with zero citations or zero references (4,417 ADS entries);
    $(ii)$ authors that have only published in Science and/or Nature (5,484 ADS entries);
    $(iii)$ entries with no authors specified (491 ADS entries);
    $(iv)$ entries with no first name for the first author (e.g. collaboration articles; 7,713 ADS entries);
    $(iv)$ entries for which first author only used initials for all publications available in the dataset (42,448 ADS entries); and
    $(iv)$ entries for which the gender of the first name of first author could not be determined (2,260 ADS entries). Note that the numbers of the ADS entries removed due to different individual reasons do not add up to the total number of removed entries due to overlaps. \\
    
    In Tables~\ref{tab:bigtab} and ~\ref{tab:bigtab2} we show twelve randomly chosen lines from our dataset as an example. We make the full final dataset available along with the non-processed data which we used\footnote{\url{http://people.phys.ethz.ch/~caplarn/GenderBias/}}.\\

In Figure~\ref{fig:sample} we present the number of papers in our sample. In the upper panel we show the number of published papers per year over time, the number of papers for which we were able to recognize the gender (the main sample we discuss in this work), and then finally the number of papers published by male and female authors. In the lower panel the same information is presented as a fraction of papers with recognized gender, male authorship or female authorship. We see that we are able to recognize gender for large fraction of papers, ranging from 60\% in the 1960s and 1970s and rising to 75\% to 80\% in the 1980s to 2010s. The fraction of recognized papers is slightly decreasing in the last few years as the fraction of authors which have published a single or few papers increases; for these authors it is less likely that the full author name is available from one of their papers. We discuss the possible influence of this effect on our results in Section \ref{sec:discussion}. We also note the slow but constant rise of the fraction of female first author papers, from less than 5\% in the 1960s to about 25\% in 2015. This trend is consistent with the overall increase in women faculty members in astronomy departments \citep{Ivie13}.

\section{Characteristics of Publications in Astronomy}    \label{sec:SampleChar}

    In this section, we highlight a few average trends concerning the characteristics of publications in astronomy. Specifically, we focus on the average number of references and citations, average seniority, and publication frequency of males and females. In Appendix \ref{app:further_sample} we also investigate who is leaving the field of astronomy and the length of papers authored by females and males.
    
\subsection{Global trends of the sample}

	Figure~\ref{fig:sample_characteristics} shows a few global average trends of publications in astronomy and highlights differences between male and female first author papers. In panel (a), we plot the average number of references in papers as a function of the publication year. We note the strong increase in the number of references per paper over time: the average number of references increased from about 10 per paper in the 1960s to $\sim60$ today, a $\sim500\%$ increase. This effect accompanies the similarly strong increase in the total number of papers published per year (see Figure~\ref{fig:sample}). The striking feature about Figure~\ref{fig:sample_characteristics}, panel (a), is the difference between the average number of references between male and female first author papers. From 1980 onward, we find a clear trend that female first author papers contain $7\pm3\%$ more references than male first author papers. A similar trend can be found for the length of papers, where female first author papers tend to be longer than male first author papers (see Appendix~\ref{app:further_sample}). \\

	Figure~\ref{fig:sample_characteristics}, panel (b), shows the average seniority of first authors in a given year. Since the seniority is determined from the first publication found in our database, we are not able to determine the seniority accurately before 1978. We reach the $>90\%$ completeness in seniority by 1978, i.e., trends before that date must be interpreted with care (see Appendix~\ref{app:completness} for more details). We find that the average seniority of $\sim7$ years in 1980 for both female and male first author papers. After that point we find that the average seniority of the male first author papers increases steadily, while the one of the female first author papers remains roughly constant.\\

	Figure~\ref{fig:sample_characteristics}, panel (c), compares the average number of citations a paper receives for males and females as a function of publication year. Up to year 2000, the average number of citations slightly increased for both male and female first author papers. The recent down-turn can be explained by the trivial effect that not enough time has passed in order to cite those papers. Overall, we find an indication that the male first author papers have on average a higher citation count than female first author papers. We will investigate this further in Sections~\ref{sec:simple_bias} and \ref{sec:ml_bias}, which contain the main part of our analysis. \\

	Finally, in Figure~\ref{fig:sample_characteristics}, panel (d), we investigate gender representation in all of the journals selected. We find that female authors tend to be underrepresented in the most prestigious journals which tend to gather most citations. Around 1980, the fraction of female first author papers was similar in all journals and amounted to $\sim10\%$. Until 2015, the fraction of female first author papers increased in all journals, but more significantly in A\&A, ApJ and MNRAS (to $\sim25\%$) than in Nature and Science (to $\sim17\%$). As an example of the difference between the journals, for papers published in the year 2000, the number of citations that papers published in Science and Nature received is two times larger than for papers published in A\&A and MNRAS, while  ApJ papers have received around 40\% more citations than those published in A\&A and MNRAS.

\subsection{Publishing frequency}   

\begin{figure}[htp]
    \centering
    \includegraphics[width=.99\linewidth]{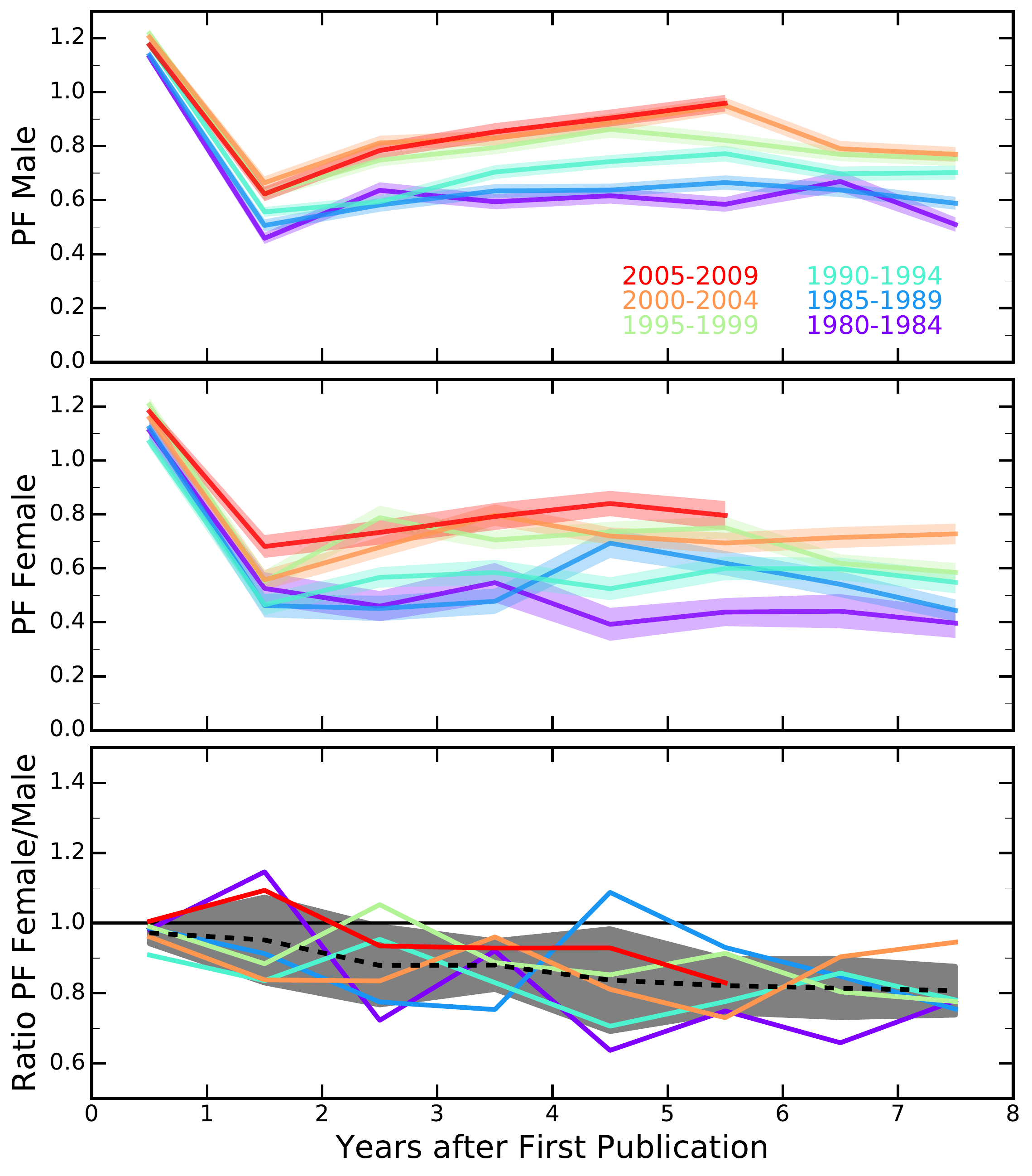}
    \caption{Publishing frequency (PF) as a function of years after the first publication (seniority). The panels show the PF for males (top), for females (middle), and the ratio of the latter and the former (bottom). Dashed line in the bottom panel denotes the mean female/male publishing frequency ratio, with shaded region denoting 1 $\sigma$ spread. The data is stacked in 5-year intervals. From the 1980s to 2015, the PF nearly doubled for both males and females. Furthermore, the PF of females decreases with respect to that of males with time after their first publication to $\sim81\%$ after 7 years.}
    \label{fig:publication_frequency}
\end{figure}
    
    In order to assess a gender dependence in productivity of publishing papers, we look into the mean number of published papers per year in seven years after their first publication (publishing frequency; PF). Here, we have excluded authors that have left the field of astronomy, i.e. we require that authors have at least one publication ten or more years after their first one. \\
    
    Figure~\ref{fig:publication_frequency} shows the PF for males and females, and the ratio of the two, stacked in 5-year intervals. In the first year, the average male and female authors sometimes publish more than one paper, leading to a PF of 1.1-1.2. The first result concerns the trend of the PF with time from 1980 to 2009: the PF increased for both males and females from 1980 to 2009 by nearly a factor of 2. At present, a typical researcher publishes nearly twice as many papers per year compared to 30 years ago. This trend is stronger for females than for males: for males the PF increased from 0.6 to 0.9 from 1980s to today, while for females the PF increased from 0.4 to 0.8. \\
        
    The bottom panel of Figure~\ref{fig:publication_frequency} shows the ratio of the male and female PF. Within two years after the first publication, the ratio is close to 1, indicating no major difference between male and female publications. However, the ratio drops steadily and reaches, seven years after the first publication, a value of 0.81, indicating that females publish $19\pm7\%$ fewer papers than males. %In Appendix~\ref{app:further_sample} we investigate the fraction of males and females leaving the field of astronomy as a function of years after their first publication.

\section{Gender difference} \label{sec:simple_bias}

\begin{figure*}[htp]
    \centering
    \includegraphics[width=.99\textwidth]{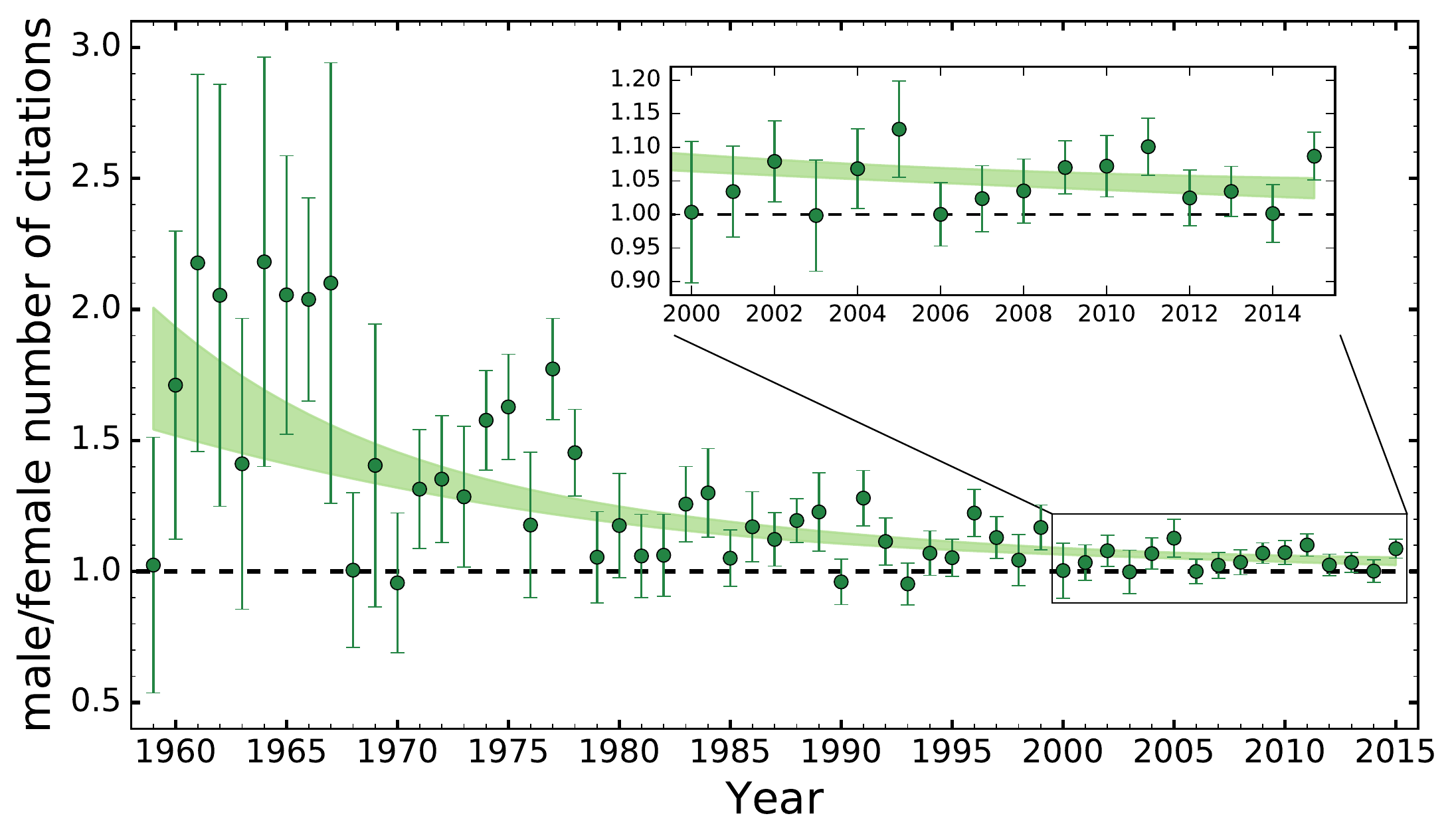}
    \caption{Gender difference: ratio of mean number of citation of male over female first author papers. The error bars are obtained by bootstrapping. The shaded green area shows 1-$\sigma$ uncertainty on the best fit (see text for details). In every year since 1960, male first author papers receive on average similar or more citations than female first author papers. This difference was higher at earlier times, although the measurement is also more uncertain. Since 1990, this difference stayed roughly constant at $\sim5\%$. The inset shows a zoom-in on the years 2000 to 2015.}
    \label{fig:MasterPlot}
\end{figure*}

	In this section we examine whether there is differences between male and female papers in terms of number of citations. As highlighted before (Section~\ref{sec:SampleChar}), male and female papers have different properties in the sample. Since the citation count is expected to correlate with certain non-gender specific properties of the papers (such as seniority and number of references), one has to be careful when interpreting the quoted difference in the number of citations. We will separate the gender bias effect from the effect caused by non-gender specific properties of the papers in Section \ref{sec:ml_bias}.  \\

	We report our results in Figure \ref{fig:MasterPlot}, which shows the mean number of citations received by male authors divided by the mean number of citations received by female authors in a given year. Errors for a given year are derived by bootstrapping. We also show the results of fitting the data with the functional form of $a_{2} e^{a_{1}(y_{t}-y)}+a_{3}$, where y is year. The best fit parameters are $a_{1}=0.06\pm0.02$, $a_{2}=1\pm0.04$, $a_{3}=0.38\pm0.24$ and $y_{t}=1974\pm12$. \\

	In the early years of our sample we see a large difference between the male and the female papers, with male authors receiving between 50\% and 100\% more citations than female authors. Of course, in this early period the errors are large due to small number of papers in total and even smaller number of female first author papers (see Figure~\ref{fig:sample}). Overall, the difference has decreased over time and appears to slowly saturate at $\sim5\%$. \\

	To quantify the difference we introduce the variable, $b_{y}$, defined as linear fit to the data presented in Figure \ref{fig:MasterPlot} after a certain year. In this work we will use the year 1985 as the cutoff year, i.e., $b_{y}$ is obtained by linearly fitting the data from 1985 to 2015. Thus, we search for $b_{y}$ that minimizes 
\begin{equation} \label{eq:bias}
\sum_{y>(y_{min}=1985)} \frac{(d_{y}-b_{y_{min}})^{2}}{\sigma^{2}_{d_{y}}},
\end{equation}
\noindent
where the $d_{y}$ is the gender difference measured in given year and $\sigma_{d_{y}}$ is estimated error of the measured gender difference. Using this definition we find the value of $b_{1985}=1.056 \pm 0.010$. Changing the cutoff year does not significantly change our results, because the fit is always dominated by the data points in the latter years due to their small errors. For example, when taking the cutoff year to be 2000, we find that $b_{2000}=1.046 \pm 0.009$.

\subsection{Controlling for seniority}

\begin{figure}[htp]
    \centering
    \includegraphics[width=.49\textwidth]{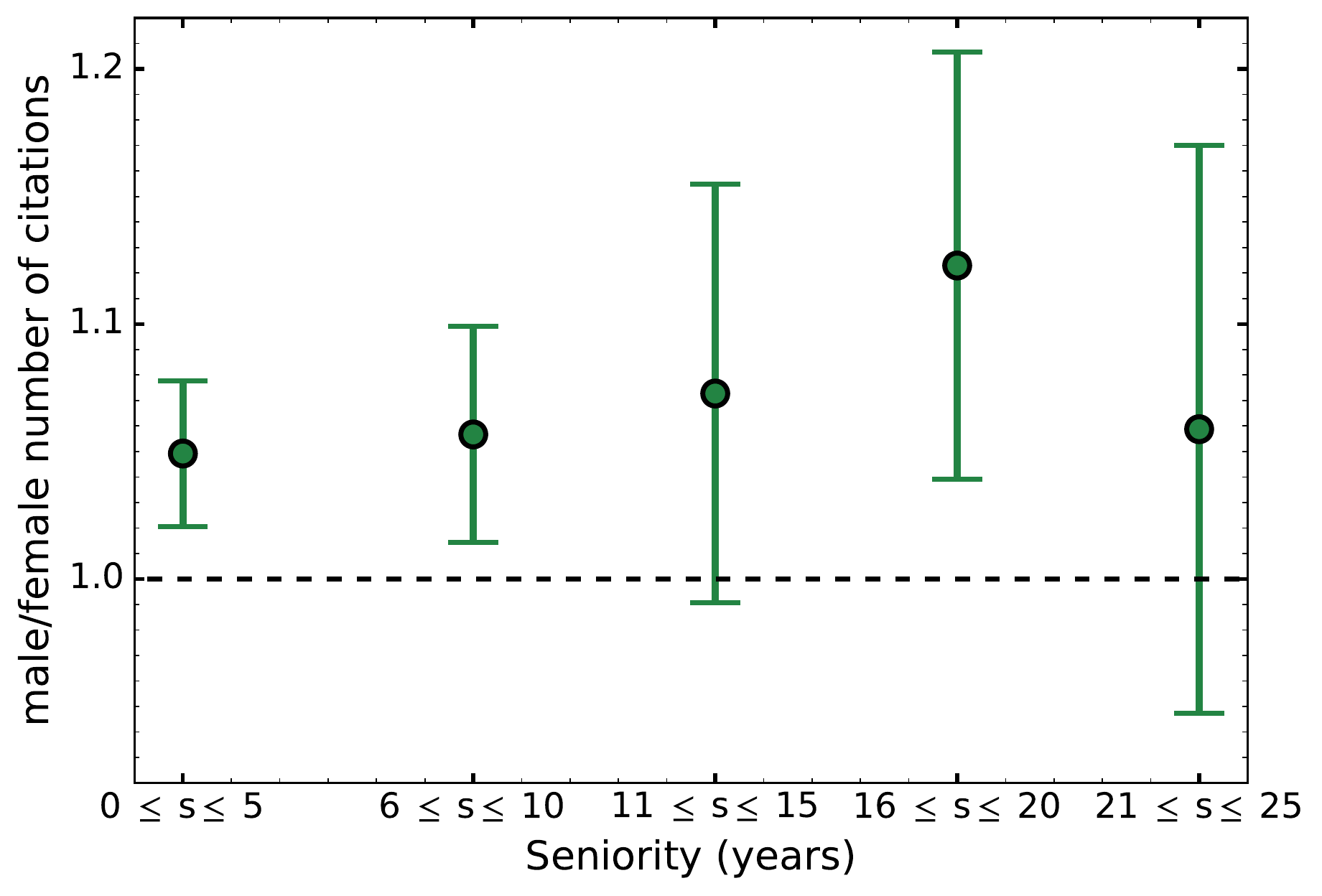}
    \caption{Difference in the number of citations for male and female first author papers, separated in 5-year wide seniority bins. We denotes the seniority with $s$ in the label. The gender difference, $b_{1985}$, is measured as described in the text. We find no significant trend of the gender difference with seniority, indicating that male first author papers receive about 5\% more citations than female first author papers at all stages of their careers.}
    \label{fig:SenPlot}
\end{figure}

	As described in Section \ref{sec:SampleChar} the male and female samples differ intrinsically. In this section, we address the question of whether the measured difference is due to different distributions of the seniority of male and female first authors. We could imagine that in a case when male first authors are on average more senior (see also Figure~\ref{fig:sample_characteristics}), they would also receive more citations since they are more established in their field. To investigate this, we split our sample into 5 year bins according to the seniority of the author and repeat the same analysis as described above. We report our results in Figure \ref{fig:SenPlot} where we show the measured gender difference, $b_{1985}$ as defined in Equation \ref{eq:bias}, for the sample which is split in this way.\\ 

	We note that the measured gender difference is still present even in the datasets in which the males and the females are matched in seniority. We see no significant trend of measured difference with seniority, indicating that male first author papers receive about 5\% more citations than female first author papers at all stages of their careers. The difference is present even in the bin with lowest seniority, i.e., when the author has just entered the field. As these samples are smaller than the full sample the errors are naturally larger, which prevents us from making any further stronger conclusions. 

\section{Gender Bias: Correcting Gender Difference for non-gender specific Properties with Machine Learning} \label{sec:ml_bias}

	The discussion above implies that it is complex to estimate the amount of gender bias given the large difference in properties of female and male first author papers. Any difference that we see could just be the consequence of the fact that female and male authors publish inherently different papers and, hence, may receive fewer citations not because of their gender, but because of some other parameter. Given that there are many possible variables influencing the citation number of the papers it is impossible to isolate or study a single variable (e.g., seniority as discussed above) to capture the full span of possibilities influencing our estimate of gender bias. Therefore, we resort to machine-learning techniques in order to correct and estimate more accurately the amount of gender bias. \\

	The main idea is to train the random-forest algorithm on the sample of male first author papers, using all the non-gender specific parameters available for the dataset. We will then use the trained algorithm to estimate the number of expected citations from the papers written by female first authors, given the properties of their papers. By comparing the predicted to the measured number of citations, we will be able to constrain the intrinsic gender bias, which is corrected for non-gender specific properties of male and female first author papers. \\ 
    
\subsection{Constructing samples and training the random-forest algorithm}

	In this analysis, we characterize the papers by using the following non-gender specific properties: seniority of the first author, number of references, number of authors, year of publication, publication journal, field of study and the region of the first author institution. We do not use the paper properties that do not span the whole dataset (e.g., number of words in a paper), as we wish to characterize the evolution of gender bias through time. Furthermore, we remove papers that do not have this information. In particular, we remove 22,685 that do not have an institution region. Importantly, our results do not change significantly if we include these papers and remove geographical information as one of the parameters in the analysis.  \\
    
	From the total male dataset we created a training and a testing subsample by randomly drawing papers. We created the testing subsample so that it contains in each year the same number of papers as the female sample. This assures that the estimates of the error in each year are comparable between the testing and female sample. \\
    
    We then searched for optimal parameters of the random forest algorithm (number of trees, minimal leaf size and number of parameters considered when splitting) in the following manner. We use the trained random-forest model on the testing subsample to generate mock number of citations and then use on them the same procedure as described in Section \ref{sec:simple_bias} to evaluate the difference between the training and the testing set. We choose the values which show no difference between the training and the testing sets (tree number=50, maximal number of features considered when splitting=80\% of available parameters, minimal leaf size=20). The code is openly available\footnote{\url{http://people.phys.ethz.ch/~caplarn/GenderBias/}}. We checked our results by running the code 40 times with different randomly selected training and testing subsamples and find that these results are robust. We use scikit-learn Python package for this analysis, but we also confirm that results are unchanged when using Wolfram Mathematica implementation of the random forest algorithm. Most important predictive features in the dataset, measured with the ``Gini importance'' estimator \citep{B84} are number of references, year of publication and journal, respectively. \\   
   
\subsection{10\% fewer citations for female first author papers}
        
\begin{figure*}[htp]
    \centering
    \includegraphics[width=.99\textwidth]{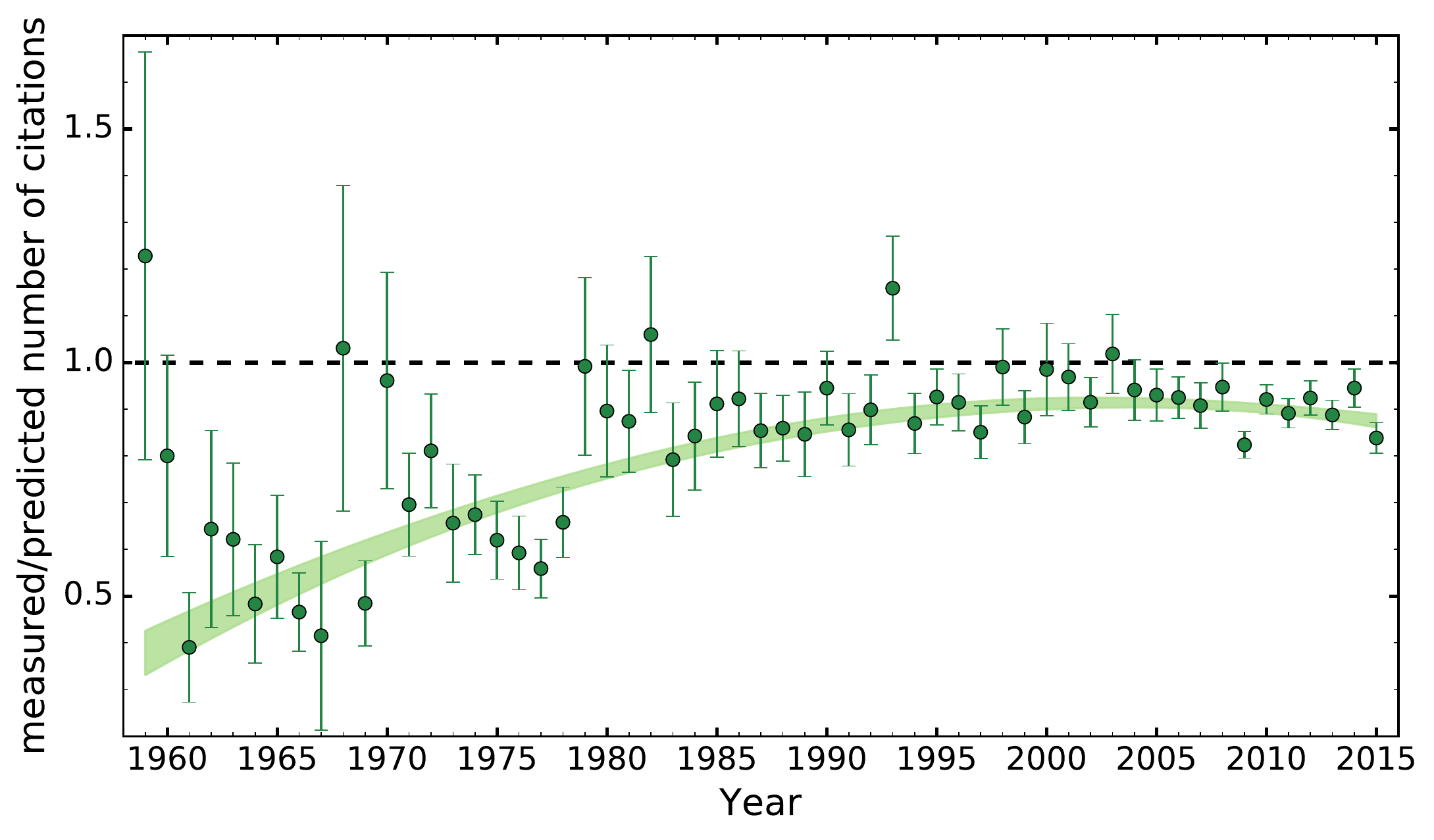}
    \caption{Gender bias: measured over predicted number of citations for papers authored by females. The error bars are obtained by bootstrapping. The shaded green area shows 1-$\sigma$ uncertainty on the best fit (see text for details). The predictions of the citation numbers are based on non-gender specific properties of papers authored by males. We measure an average intrinsic bias of about 10\%, implying that females systematically receive around 10\% fewer citations than what would be expected if they were males, given the properties of their papers.}
    \label{fig:FigFemFem}
\end{figure*}

    Having verified that our trained algorithm is able to accurately predict the number of citations based on a several non-gender specific parameters, i.e. does not find a difference between the training and testing samples which were both drawn from the sample of male first author papers, we used the same algorithm on the female sample. In Figure \ref{fig:FigFemFem} we show the ratio of the measured number of citations that female authors have received to the number of citations that would be expected from our analysis. We find that papers with female authors systematically receive fewer citations than what would be expected given the other, non-gender specific properties of their papers. We also show the results of fitting the data with the functional form of $b_{1} (y_{t}-y)+b_{2}(y_{t}-y)^{2}+b_{3}$, for which the best fit parameters are $b_{1}=-0.03\pm0.003$, $b_{2}=-0.00026\pm0.00004$, $b_{3}=0.04\pm0.11$ and $y_{t}=1945 \pm5$. We define the quantity $b_{\rm ff^{'}}$, characterizing this bias between the simulated female sample ($f^{'}$) and the actual female sample ($f$), and measured by fitting the data presented in Figure \ref{fig:FigFemFem} from the year 1985, with the same procedure as presented in Section \ref{sec:simple_bias}. We measure this bias to be $b_{\rm ff^{'}}=0.896 \pm 0.009$, i.e. we find that females systematically receive around 10\% fewer citations than that what would be expected given the properties of their papers. \\
      
	To check the consistency of our results presented here (bias that amounts to $10\%$) and the ones in Section~\ref{sec:simple_bias} (uncorrected gender difference that amounts to $6\%$), we replace the measured number of citations a female first author paper receives with the predicted number of citations. With this experiment we consider what would be the difference in number of citations if there was no intrinsic bias between male and female first author papers. We measure this value to be $b_{\rm mf^{'}} = 0.958 \pm 0.008$. In order words, if there was no intrinsic bias between the male and female authors we would expect that male authors in our sample should receive 4$\%$ fewer citations than the female authors, purely from the differences in the properties of their papers. As shown above, we detect that actually male authors receive around $6\%$ more citations (gender difference in \ref{sec:simple_bias}). Hence, these two effects together add up to the $10\%$ difference that we measure between the expected and the measured value of citations received by the female authors. \\

\section{Self-citation} \label{sec:self_citation}

\begin{figure}[htp]
    \centering
    \includegraphics[width=.49\textwidth]{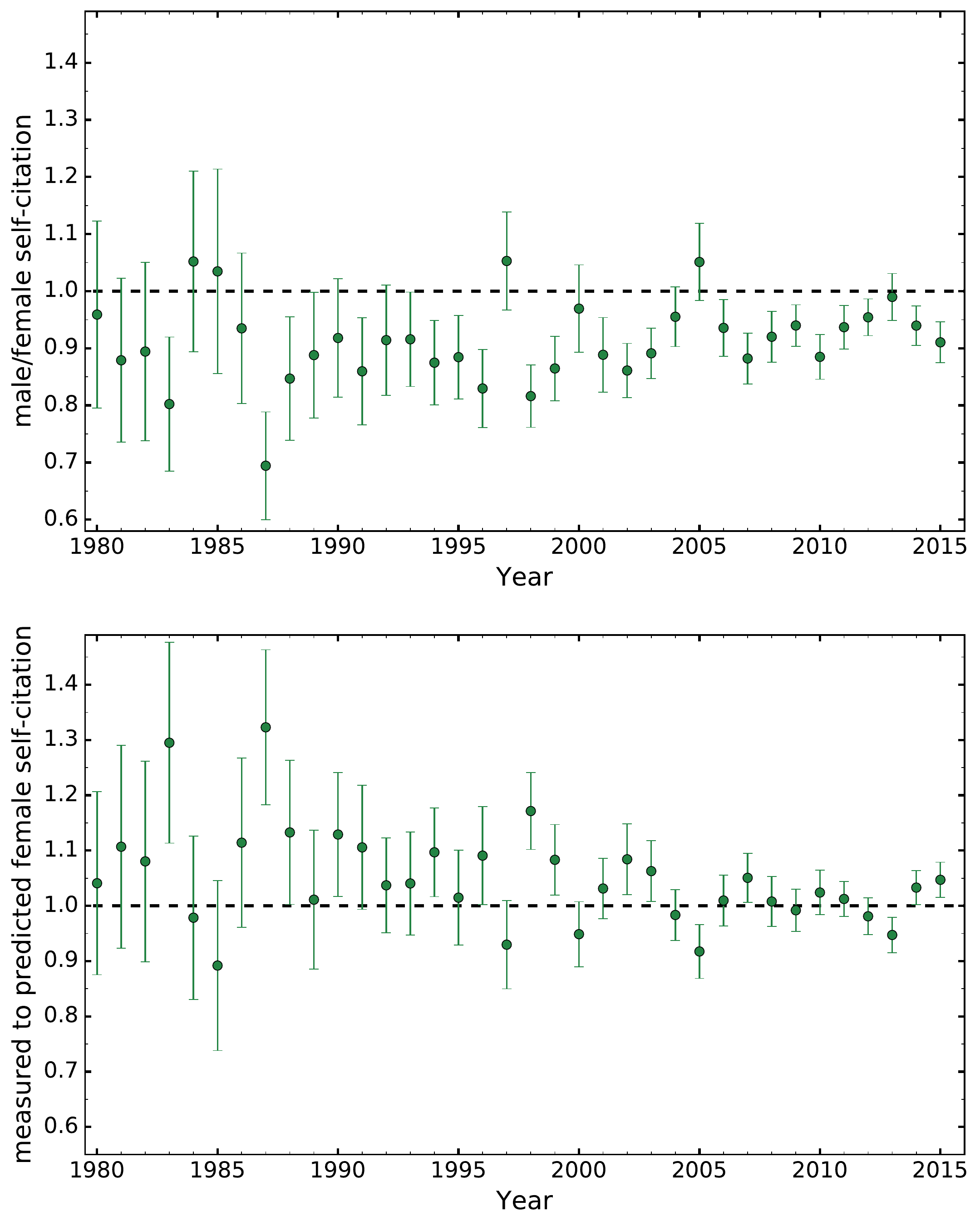}
    \caption{Self-citing tendency of males and females. The upper panel shows the ratio of the fraction of male and female first author papers that have cited their previous paper. We see that males tend to consistently self-cite less than their female colleagues. However, the lower panel shows the same, but corrected for non-gender specific properties of the papers, indicating that there is no inherent difference in the propensity for self-citation between the male and female authors when all of the other parameters are accounted for. \\}
    \label{fig:SelfCite}
\end{figure}

	There are many possible reasons to explain the aforementioned difference in the number of citations of female and male first author papers. One reason could be that males tend to cite themselves more often compared to females. This directly increases the number of citations, as well as the visibility of the paper, leading to more citations in the future. Therefore, we investigate the self-citing tendency of male and female authors in this section. \\
    
    Precisely defining ``self-citations'' is of crucial importance. During our initial analysis it became obvious that the result differs significantly depending on the definition used. This is due to the obvious correlation between the number of self-citations and the number of papers previously published by a given author. Given that the sample of males tends to have more papers this immediately leads to the conclusion, when not controlled for this correlation, that males tend to self-cite more. We believe that this is the main driver for conclusion stated in the paper by \cite{K16} who found when analyzing the whole JSTOR database of scientific papers that men cite their own papers 56\% more than women do. The fact that authors with more published papers also have more self-citations is expected, and it is hard to define an objective criteria that differentiates justified self-citations and ``unnecessary" ones. We tried various definitions and concluded that any definition that uses total number of self-citations or total number of published papers by an author carries this bias to a certain level. \\

    Therefore, we use a definition that is explicitly not dependent on the total number of self-citations of an author: for each paper in our database we check if the paper in question has cited the paper published immediately prior to it by the same first author. Therefore, for each paper in our database, we assign a value of 1 if the paper has cited the the most recently published paper by the same first author, 0 if it has not. We exclude all papers that are the first publications from our analysis. We expect our results to be skewed for authors working in multiple fields, but we do not expect a large difference between genders in this aspect. \\
    
	We proceed with our analysis in the same fashion as described in Sections \ref{sec:SampleChar} and \ref{sec:simple_bias}. For this analysis, the most important predictive features in the dataset are number of references, seniority and year of publication. Our results are illustrated in Figure \ref{fig:SelfCite}. We only show results from the year 1979 to 2015 because the low number of self-citations makes it impossible to create a meaningful comparison between the genders before 1979. In the upper panel, we plot the ratio of the fraction of male to female first author papers that have cited their previous paper. We see that males tend to consistently self-cite less than their female colleagues. Using the linear fit to the data, in the same manner as described in Equation \ref{eq:bias}, we find that the self-citation ratio (sc) between males and females after 1985 is $sc_{\rm mf,1985}=0.91 \pm 0.02$. \\

	We then follow up this results with the same type of random-forest analysis as described in Section \ref{sec:ml_bias}, but now we train the algorithm to predict the probability of self-citing the last work of the author of the paper instead. We verify, using multiple training and testing sets derived from the male sample, that the algorithm is successful in reproducing the input values and does not introduce biases in our measurement. In the lower panel of Figure \ref{fig:SelfCite} we show the result of the random-forest procedure to predict the number of self-citations by female authors and compare it to the number of measured number of self-citations. We see that there is little difference, i.e., the self-citation rate by female authors is the very similar to what we would expect if papers with the same properties were written by the male authors. A formal fit to the ratio of self-citation between the simulated female dataset and the measured female dataset is $sc_{\rm f^{'}f,1985}=1.015 \pm 0.011$. We note that in this case the choice of the starting year for the fit again makes a small difference to the final result; for instance setting starting year for the fit at 2000 yields the result of $sc_{\rm f^{'}f,2000}=1.004 \pm 0.011$. We conclude that the perceived difference in the upper panel of Figure \ref{fig:SelfCite} is fully explained by the different properties of the papers authored by male and female authors. We find only minimal inherent difference in the propensity for self-citation between male and female authors when all of the other parameters are accounted for and hence we conclude that this is not the main driver of the gender bias in the number of citations. \\
     %   \jwc{Again, what are the factors that most strongly predict self-citation?}

\section{Discussion} \label{sec:discussion}

	Although we took maximal care to avoid any biases in our own analysis, some caveats remain. \\

    Gender identification is of crucial importance for our analysis. As discussed in Section \ref{sec:data} we gather the data from first names of authors and run the name through multiple algorithms to determine the gender. This is not possible if the author is only using initials throughout their publication history. Even if there is no bias between males and females of using only their initials, we will tend to miss females because they are likely to publish less frequently and to be younger (see Figures \ref{fig:sample} and \ref{fig:sample_characteristics}). Both of these effects would bias us so that we would only be able to recognize gender of more established female authors in the field. The fact that the female authors in our sample tend to have more references (see Section~\ref{sec:SampleChar}) and longer papers (see Appendix~\ref{app:further_sample}) could be, at least partially, a manifestation of this effect. This potentially also contributes to the observation that we would expect that the female authors should receive around 4\% more citations than males in our sample (see Section~\ref{sec:ml_bias}). Additionally, this recognition problem is possibly aggravated because some females change their last names due to marriage. This effect could also lead to underestimation of seniority for some female authors as we misidentify established authors as newly arriving in the field (see Section~\ref{sec:SampleChar}). We note that fully accounting for these effects would probably increase the observed difference in citation counts between females and males in astronomy. \\
    
    Because our name classification mechanism is mostly based on data sources in Europe and North America, this means we are less likely to recognize the gender of names from different cultures. This becomes especially apparent in later years with a more globalized astronomy community. We do not expect this to create any strong effect in our analysis as we have checked that the gender bias is largely independent of the region where the host institution is based.  \\
    
    Of course we cannot claim that we have actually measured gender bias. One could imagine numerous other parameters that should be considered and matched before such a conclusion could be drawn. Our results therefore should be taken with care. It is our best effort based on all of the available data that we could acquire. We encourage the community to work on and/or enhance our dataset for further analysis. \\

\section{Conclusions} \label{sec:conclusion}

The main goal of this paper is to quantify the gender bias in astronomy. We define ``gender bias'' as the difference in the citation counts between female and male first author papers with matched non-gender properties. We assembled information about all papers published in A\&A, MNRAS, ApJ, and about all of astronomical papers published in Science and Nature from 1950 to 2015. In total, we have analyzed over 200,000 papers. Using the gender recognition algorithm, we assigned gender to the first author of every paper where this was feasible (about 70\% of all papers). For the majority of the remaining papers we were not able to deduce the gender of the first author because the author only used initials throughout their publishing career. Our main conclusions are as follows: 

\begin{itemize}

\item Female participation has been consistently rising over time. Females authored around 25\% of the papers in the last few years, but this rise has been the slowest in the most prestigious journals, such as Nature and Science, where the fraction amounts to only 17\%. 

\item By simply measuring the difference between the number of citations received by male and female authors in our sample we find a clear $5.6\pm1.0\%$ difference in favor of male authors, when measured from the year 1985 onward. This gender difference does not change significantly when choosing a later year for the measurement as the difference is decreasing very slowly or not at all. 

\item We estimate gender bias by using machine-learning techniques to control for differences between the male and female first author papers. We find that females receive $10.4\pm0.9\%$ fewer citations than what would be expected if the papers with the same characteristics were written by the male authors. This is consistent with our finding that if gender bias did not exist, we would expect males in our sample to receive 4.2$\pm0.8\%$ less citations than females. 

\item Using the probability of an author having self-cited their previous paper as a self-citation metric, we find that females in our sample are $9\pm2\%$ more likely to cite their previous work. When using machine-learning techniques to control for differences between the male and female samples we find no significant intrinsic differences in propensity of male and female authors to cite themselves. 

\end{itemize}

Our conclusions are limited by our inability to determine the gender for all of the authors of the papers. We believe that this effect would probably act in a manner to further strengthen our conclusion about the existence of gender bias in astronomy. \\

We make our dataset publicly available and invite further research on this topic.

\vspace{0.4 cm}

\acknowledgments

We would like to thank Joanna Woo for giving detailed comments on the manuscript. We acknowledge the stimulating comments given to us by Meg Urry, Renate Schubert, Raffaella Marino, Benny Trakhtenbrot, Izabela Moise and Evangelos Pournaras. We thank Amanda Bluck for proof reading the manuscript. We acknowledge support from the Swiss National Science Foundation. This research made use of NASAs Astrophysics Data System (ADS), the arXiv.org preprint server, and the Python plotting library \texttt{matplotlib} \citep{hunter07}. \\

\appendix

\renewcommand\thefigure{\thesection.\arabic{figure}}    
\setcounter{figure}{0}

\section{Further Characteristics of our Sample} \label{app:further_sample}

Clearly, the data presented in this work is very rich and many different aspects can be investigated. We present here some further details. 

\subsection{Leaving the field} 

\begin{figure}[htp]
    \centering
    \includegraphics[width=0.99\linewidth]{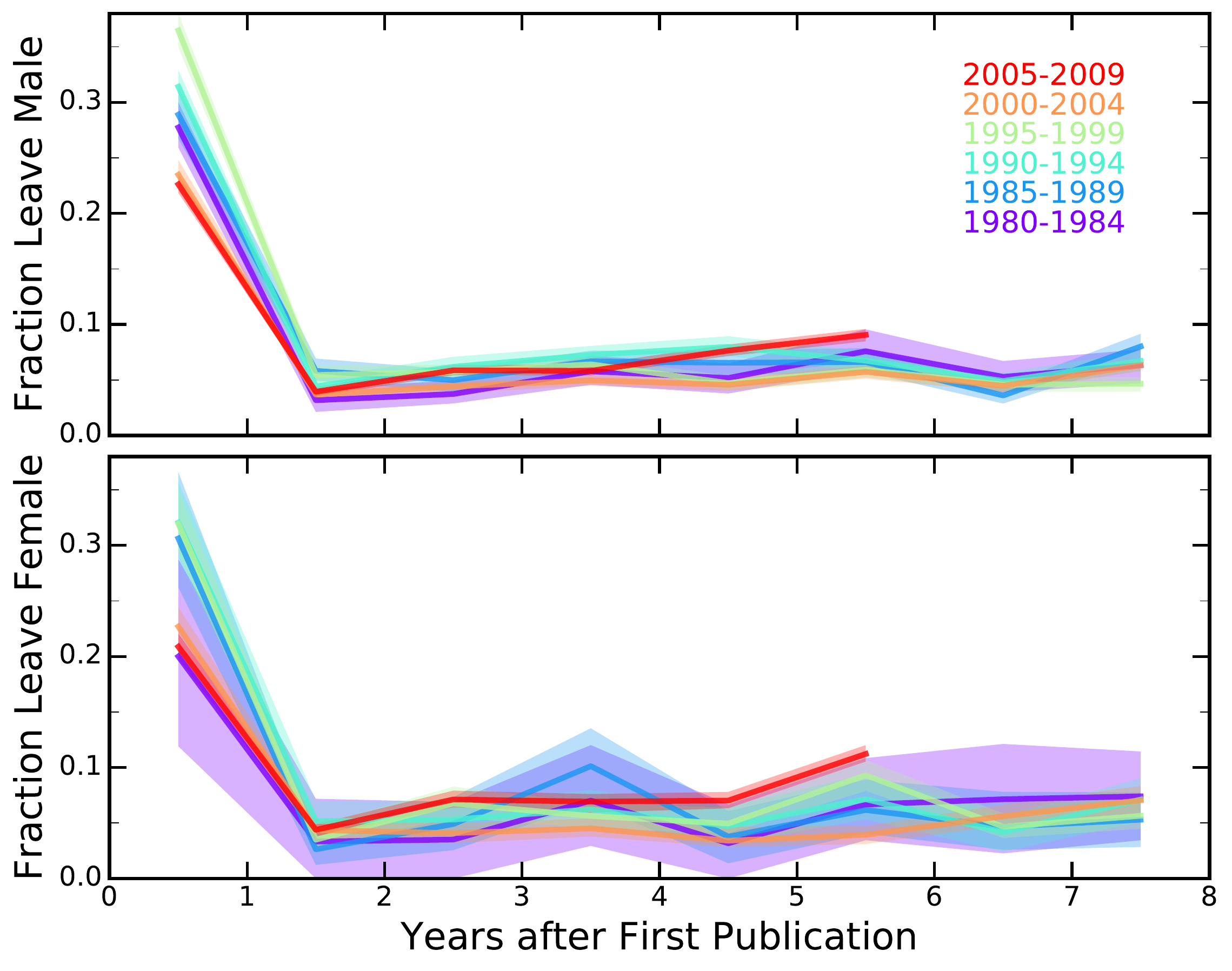}
    \caption{Fraction of males (top panel) and females (bottom panel) that leave the field as a function of years after their first publication. The data is stacked in 5-year intervals. }
    \label{fig:leave_field}
\end{figure}

	An interesting question is whether females or males are more prone to leave the field of astronomy. We address this question by looking at the publication pattern of individual male and female authors. Specifically, we measure the fraction of authors that have published their final paper (i.e., have not published another paper in the following years in of the five journals considered here) as a first author in a given year. This is done out to 8 years after their first publication. 

	Figure~\ref{fig:leave_field} shows the fraction of male and female authors that have left the field as a function of years after their first publication (seniority). We have stacked the data in 5-year intervals. Between 20-30\% of both males and females do not publish any further papers after their papers in the first year. This drops significantly to about 5\% in the second year, and stays thereafter between 5-10\%. We do not find any difference between males and females nor with time (from 1980 to today), within the measurement errors.

\subsection{Length of papers } 

\begin{figure}[htp]
    \centering
    \includegraphics[width=0.99\linewidth]{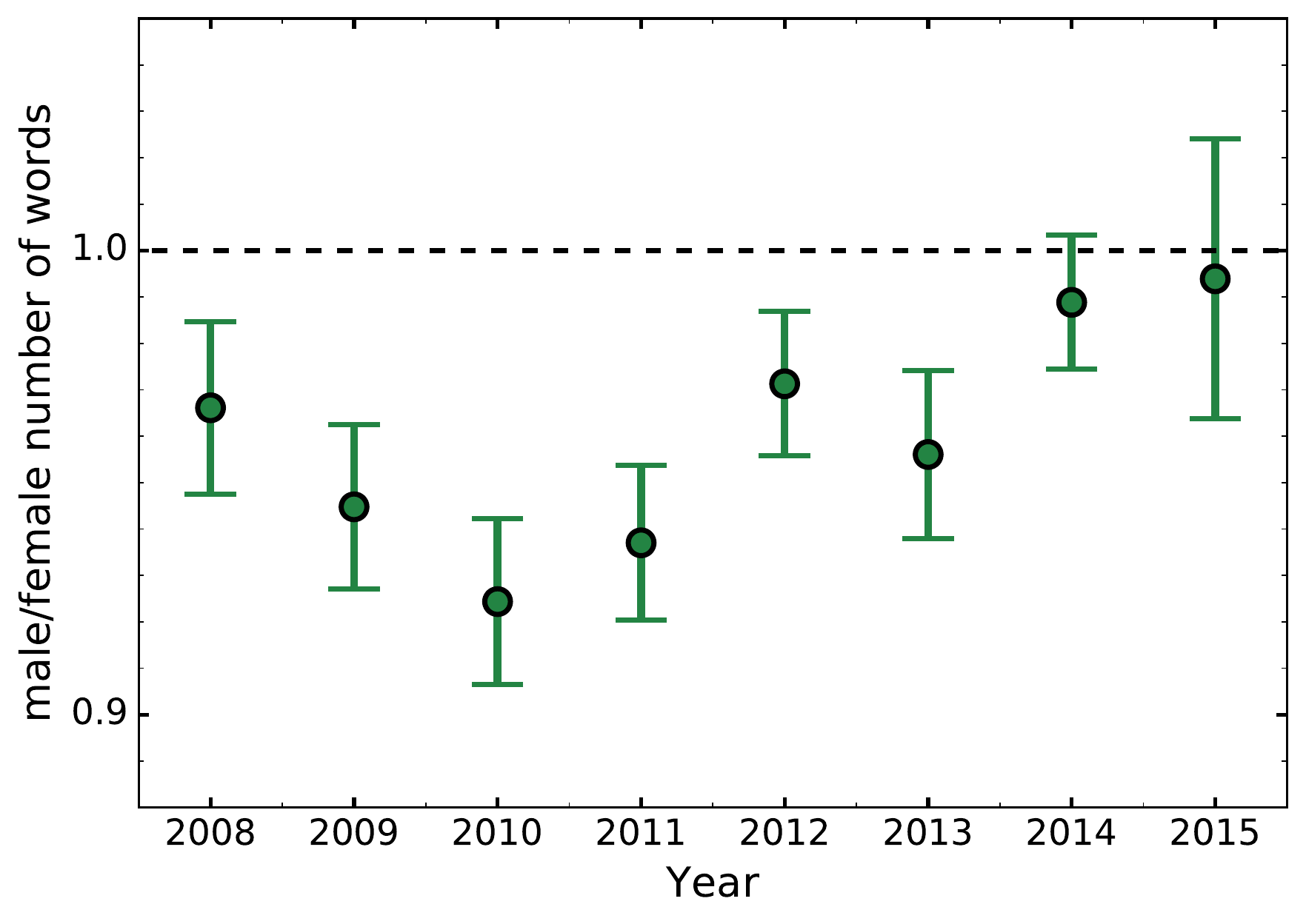}
    \caption{ Mean ratio of the length of the male first author papers to the length of the female first author papers. Male authored papers in our sample tend to be slightly shorter. }
    \label{fig:length}
\end{figure}

	Figure \ref{fig:length} shows the ratio of the length of the papers authored by males to the length of the papers authored by the females. The data for this analysis is available only from the year 2007 as this is the year from which the *.tex files of the papers are easily accessible via the S3 Amazon server. We see that papers written by the male authors in our sample tend to be around 5\% shorter than papers written by the female authors. This is consistent with our finding that the number of references in the female papers is larger than the number of references in the male papers, which is the effect which we observe across all of the years in the sample. As elaborated in Discussion (Section \ref{sec:discussion}) we can not distinguish if this is the real effect or if we are somehow biased against recognizing gender of the the female authors which tend to write shorter papers.

\section{Completeness of Seniority} \label{app:completness}

\begin{figure}[htp]
    \centering
    \includegraphics[width=0.99\linewidth]{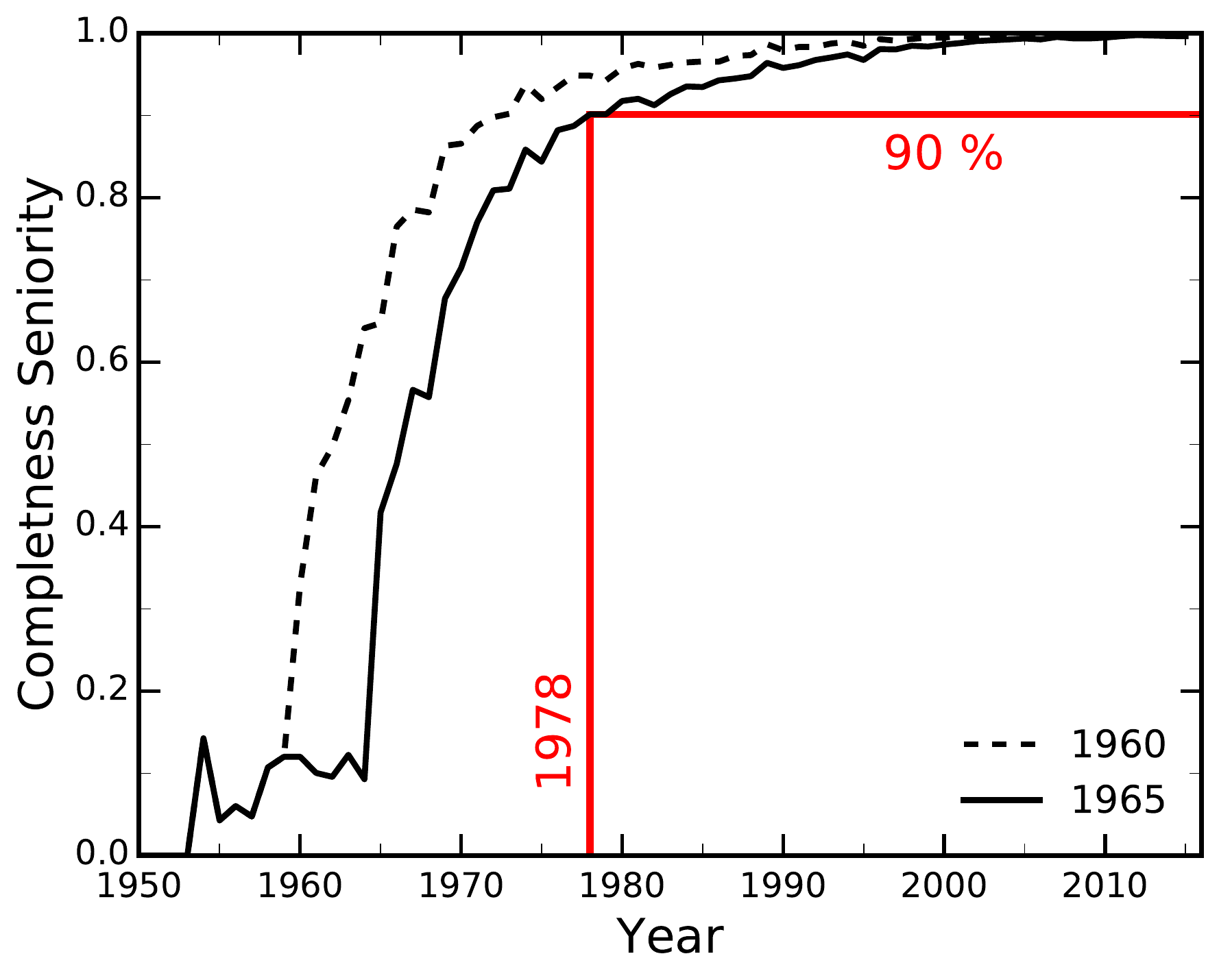}
    \caption{Completeness of seniority. The solid and dashed lines show the fraction of papers which the first author's first paper has been published before 1965 and 1960, respectively. We reach the 90\% completeness limit in 1978. }
    \label{fig:completness}
\end{figure}

	We define seniority as the number of years since the author's initial first author publication. Since only papers after 1950 are included in our analysis, there is the possibility that we do not determine the seniority accurately at early times. In Figure~\ref{fig:completness} we show the fraction of papers with first authors who have published their first paper before 1960 and 1965, respectively. We find that in 1978 90\% of all papers have a seniority after 1965, i.e., from the year 1978 we are complete at the least 90\% level. 
    
\end{document}